\documentstyle[aps,epsfig]{revtex}
\begin{document}
\title{$J/\Psi$ suppression in an expanding equilibrium hadron gas
\thanks{ Work partially supported by the Polish Committee for Scientific Research
under contract KBN - 2 P03B 030 18}}
\author{Dariusz Prorok and Ludwik Turko}
\address{Institute of Theoretical Physics, University of
Wroc{\l}aw,\\ Pl.Maksa Borna 9, 50-204  Wroc{\l}aw, Poland}
\date{February 17, 2000}
\maketitle
\begin{abstract}
We consider an ideal gas of massive hadrons in thermal and
chemical equilibrium. The finite-size gas expands longitudinally
in accordance with Bjorken law. Also the transverse expansion in a
form of the rarefaction wave is considered. We show that $J/\Psi$
suppression in such an environment, when combined with the
disintegration in nuclear matter, agrees qualitatively well with
NA38 and NA50 data.
\end{abstract}

\section {Introduction }

Since the paper of Matsui and Satz \cite{MatSatz} there is a
steady interest in the problem of $J/\Psi$ suppression in a
heavy-ion collision. The question is if this suppression can be
treated as a signature for a quark-gluon plasma or if it can be
explained by $J/\Psi$ absorption in a hadron gas which appears in
the central rapidity region (CRR) of the collision
\cite{FtLi,Vogt,GavGyu,Blaiz}.

In the following paper we shall continue our previous
investigations \cite{prtu1} into the problem of $J/\Psi$
suppression observed in a heavy-ion collision (for experimental
data see e.g. \cite{Abr} and references therein). Now, we shall
focus on the dependence of the suppression on the initial energy
density reached in the CRR.

In our model, $J/\Psi$ suppression is the result of a $c\bar c$
state absorption in a dense hadronic matter through interactions
of the type

\begin{equation}
c\bar c+h \longrightarrow D+\bar D+X\\ ,  \label{1}
\end{equation}

%%%%%%%%%%%%%1111

where $h$ denotes a hadron, $D$ is a charm meson and $X$ means a
particle which is necessary to conserve the charge, baryon number
or strangeness. The hadronic matter is in a state of an ideal gas
of massive hadrons in thermal and chemical equilibrium and
consists of all species up to $\Omega^{-}$ baryon. Time evolution
is given here by conservation laws combined with assumptions about
the space-time structure of the system. A corresponding equation
of state of the ideal gas makes then possible to express gas
parameters such as temperature and chemical potentials as
functions of time.

\begin{figure}
\begin{center}{
{\epsfig{file=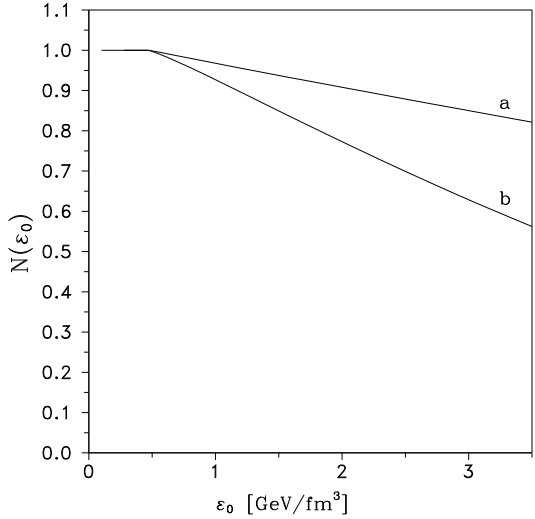,width=7cm}}
}\end{center}
\caption{comparison of suppression of the pure $J/\Psi$'s in the
cooling hadron gas for two values of the power $a$ in
approximation $T(t)\protect\cong T_0 \cdot t^{-a}$: a) $a= {1\over
3}$ ; b) $a = {1\over 5.6}$}
\label{Fig.1.}%1
\end{figure}

An ideal gas of real hadrons has a very interesting feature: it
cools much slower than a pion gas when expands longitudinally. We
have checked numerically that for the initial energy densities
$\epsilon_0$ corresponding to initial temperatures $T_{0}$ of the
order of 200 MeV and for the freeze-out $T_{f.o.}$ not lower than
about 100 MeV, the time dependence of the temperature of the
expanding gas still keeps the well-known form $T(t)=T_{0} \cdot
t^{-a}$ (we put $t_{0}=1 fm$). Only the exponent $a$ changes from
${1 \over 3}$ for massless pions to the values ${1 \over
{5.6}}$-${1 \over {5.3}}$ for massive realistic hadrons. As a
result, the time of the freeze-out $t_{f.o}$ is much greater for
the hadron gas than for the pion one. For instance, when we take
$T_{0}=200 MeV$ and $T_{f.o.}=140 MeV$ we obtain $t_{f.o.}={7.37}
fm$ ($a={1 \over {5.6}}$) for the hadron gas and $t_{f.o.}={2.9}
fm$ ($a={1 \over 3}$) for the pion gas. The lower $T_{f.o.}$, the
stronger difference. For $T_{0}=200 MeV$ and $T_{f.o.}=100 MeV$ we
have $t_{f.o.}={48.5} fm$ ($a={1 \over {5.6}}$) for the hadron gas
and $t_{f.o.}={8} fm$ ($a={1 \over 3}$) for the pion gas. This has
a direct consequence for $J/\Psi$ suppression: the longer the
system lasts, the deeper suppression causes (see Fig.~
\ref{Fig.1.}).

We are going to calculate a survival factor for $J/\Psi$ when new,
more realistic conditions are taken into account. We consider a
hadronic gas which is produced in the CRR region. This gas expands
both longitudinally and transversely. The longitudinal expansion
is a traditional adiabatic hydrodynamical evolution \cite{Bjor},
the transverse expansion is considered as the rarefaction wave. An
initial energy density depends now on the impact parameter $b$ and
on the geometry of the collision.

\section { The expanding hadron gas }
\label{hadgas}

For an ideal hadron gas in thermal and chemical equilibrium, which
consists of $l$ species of particles, energy density $\epsilon$,
baryon number density $n_{B}$, strangeness density $n_{S}$ and
entropy density $s$ read ($\hbar=c=1$ always)

\begin{mathletters}
\label{eqstate}
\begin{equation}
\epsilon = { 1 \over {2\pi^{2}}} \sum_{i=1}^{l} (2s_{i}+1)
\int_{0}^{\infty} { { dpp^{2}E_{i} } \over { \exp \left\{ {{ E_{i}
- \mu_{i} } \over T} \right\} + g_{i} } } \ , \label{2a}
\end{equation}

\begin{equation}
n_{B}={ 1 \over {2\pi^{2}}} \sum_{i=1}^{l} (2s_{i}+1)
\int_{0}^{\infty} { { dpp^{2}B_{i} } \over { \exp \left\{ {{ E_{i}
- \mu_{i} } \over T} \right\} + g_{i} } } \ , \label{2b}
\end{equation}

\begin{equation}
n_{S}={1 \over {2\pi^{2}}} \sum_{i=1}^{l} (2s_{i}+1)
\int_{0}^{\infty} { { dpp^{2}S_{i} } \over { \exp \left\{ {{ E_{i}
- \mu_{i} } \over T} \right\} + g_{i} } } \ , \label{2c}
\end{equation}

\begin{equation}
s={1 \over {6\pi^{2}T^{2}} } \sum_{i=1}^{l} (2s_{i}+1)
\int_{0}^{\infty} { {dpp^{4}} \over { E_{i} } } { { (E_{i} -
\mu_{i}) \exp \left\{ {{ E_{i} - \mu_{i} } \over T} \right\} }
\over { \left( \exp \left\{ {{ E_{i} - \mu_{i} } \over T} \right\}
+ g_{i} \right)^{2} } }\ , \label{2d}
\end{equation}
\end{mathletters}

where $E_{i}= ( m_{i}^{2} + p^{2} )^{1/2}$ and $m_{i}$, $B_{i}$,
$S_{i}$, $\mu_{i}$, $s_{i}$ and $g_{i}$ are the mass, baryon
number, strangeness, chemical potential, spin and a statistical
factor of specie $i$ respectively (we treat an antiparticle as a
different specie). And $\mu_{i} = B_{i}\mu_{B} + S_{i}\mu_{S}$,
where $\mu_{B}$ and $\mu_{S}$ are overall baryon number and
strangeness chemical potentials respectively.

We shall work here within the usual timetable of the events in the
CRR of a given ion collision (for more details see e.g.
\cite{Blaiz}). We fix $t=0$ at the moment of the maximal overlap
of the nuclei. After half of the time the nuclei need to cross
each other, matter appears in the CRR. We assume that soon
thereafter matter thermalizes and this moment, $t_{0}$, is
estimated at about 1 fm \cite{Blaiz,Bjor}. Then matter starts to
expand and cool and after reaching the freeze-out temperature it
is no longer a thermodynamical system. We denote this moment as
$t_{f.o.}$. As we have already mentioned in the introduction, this
matter is the hadron gas, which consists of all hadrons up to
$\Omega^{-}$ baryon. The expansion proceeds according to the
relativistic hydrodynamics equations and for the longitudinal
component we have the following solution (for details see e.g.
\cite{Bjor,Cley})

\begin{equation}
s(t)= { {s_{0}t_{0}} \over t }
\;,\;\;\;\;\;\;\;\;\; n_{B}(t)= { {n_{B}^{0}t_{0}} \over t } \ ,
\label{3}
\end{equation}

where $s_{0}$ and $n_{B}^{0}$ are initial densities of the entropy
and the baryon number respectively. The superimposed transverse
expansion has the form of the rarefaction wave moving radially
inward with a sound velocity $c_{s}$ \cite{Bjor,Baym}.

To obtain the time dependence of temperature and baryon number and
strangeness chemical potentials one has to solve numerically
equations (\ref{2b} - \ref{2d})  with $s$, $n_{B}$ and $n_{S}$
given as time dependent quantities. For $s(t)$, $n_{B}(t)$ we have
expressions (\ref{3})  and $n_{S}=0$ since we put the overall
strangeness equal to zero during all the evolution (for more
details see \cite{prtu2}).

The sound velocity squared is given by  $c_{s}^{2}= { {\partial P}
\over {\partial \epsilon} }$ and can be evaluated numerically
\cite{prtu2,prtu3}.

\section { J/$\Psi$ absorption in hadronic matter }

In a high energy heavy-ion collision, charmonium states are
produced mainly through gluon fusion and it takes place during the
overlap of colliding nuclei. For the purpose of our model, we
shall assume that all $c\bar{c}$ pairs are created at the moment
$t=0$. Before the fusion, gluons can suffer multiple elastic
scattering on nucleons and gain some additional transverse
momentum in this way \cite{Huf,Gav,BlaizOll}. This manifests for
instance in the observed broadening of the $p_{T}$ distribution of
$J/\Psi$ \cite{Bad}. Following \cite{Gup}, we express this effect
by the transverse momentum distribution of the charmonium states
of the form

\begin{equation} g(p_{T},\epsilon_{0}) = { {2p_{T}}
\over {\langle p_{T}^{2}\rangle_{J/\Psi}^{AB}(\epsilon_{0})} }
\cdot \exp \left\{- { { p_{T}^{2}} \over {\langle
p_{T}^{2}\rangle_{J/\Psi}^{AB}(\epsilon_{0})} } \right\}\ ,
\label{4}
\end{equation}

where $\langle p_{T}^{2}\rangle_{J/\Psi}^{AB}(\epsilon_{0})$ is
the mean squared transverse momentum of $J/\Psi$ gained in an A-B
collision with the initial energy density $\epsilon_{0}$. The
momentum can be expressed as (for details see \cite{Bagl})

\begin{equation}
\langle p_{T}^{2}\rangle_{J/\Psi}^{AB}(\epsilon)=\langle
p_{T}^{2}\rangle_{J/\Psi}^{pp} + K \cdot \epsilon \ , \label{5}
\end{equation}

with $K=0.27 fm^{3} \cdot GeV$ and $\langle
p_{T}^{2}\rangle_{J/\Psi}^{pp}=1.24 GeV^{2}$ taken from a fit to
the $J/\Psi$ data of NA38 Collaboration \cite{Bagl}. The
expression in (\ref{4})  is normalized to unity and is treated as
the initial momentum distribution of charmonium states here.

For the simplicity of our model, we shall assume that all
charmonium states are completely formed and can be absorbed by the
constituents of a surrounding medium from the moment of creation.
It means that we neglect a whole complex process of $J/\Psi$
formation as presented in \cite{Khar,Satz}. The main feature of
the above-mentioned process is that, soon after the moment of
production, the $c\bar{c}$ pair binds a soft gluon and creates a
pre-resonance $c\bar{c}-g$ state, from which, after a time of the
order of 0.3 fm, a physical charmonium state is formed. This means
that the possible nuclear absorption of charmonium is, in fact,
the absorption of the $c\bar{c}-g$ state. But the latest has the
cross-section $\sigma_{abs}=7.3 mb$, which is much higher than
$J/\Psi-Nucleon$ absorption cross-section $\sigma_{\psi N} \cong
3-5 mb$ obtained from p-A data \cite{Bad,Gers,GavVo}. This
justifies our assumption: taking into account $c\bar{c}-g$
absorption instead of charmonium disintegration in the nuclear
matter would only strengthen $J/\Psi$ suppression.

According to the above assumption, charmonium states can be
absorbed first in the nuclear matter and soon later, when the
matter appears in the CRR, in the hadron gas. Since these two
processes are separated in time, $J/\Psi$ survival factor for a
heavy-ion collision with the initial energy density
$\epsilon_{0}$, may be written in the form

\begin{equation}
{\cal N}(\epsilon_{0}) = {\cal N}_{n.m.}(\epsilon_{0}) \cdot {\cal
N}_{h.g.}(\epsilon_{0})\ , \label{6}
\end{equation}

where ${\cal N}_{n.m.}(\epsilon_{0})$ and ${\cal
N}_{h.g.}(\epsilon_{0})$ are $J/\Psi$ survival factors in the
nuclear matter and the hadron gas, respectively. For ${\cal
N}_{n.m.}(\epsilon_{0})$ we have the usual approximation
\cite{GavVo,GerH,Gavhep}

\begin{equation}
{\cal }_{n.m.}(\epsilon_{0}) \cong \exp \left\{ -\sigma_{\psi N}
\rho_{0} L \right\}\ , \label{7}
\end{equation}

where $\rho_{0}$ is the nuclear matter density and $L$ the mean
path length of the $J/\Psi$ through the colliding nuclei. For the
last quantity, we use the expression given in \cite{Gavhep}:
\begin{equation}
 L(b) =
{1 \over {2 \rho_{0} T_{AB}} } \int d^{2}\vec{s}\; T_{A}(\vec{s})
T_{B}(\vec{s} - \vec{b}) \left[ T_{A}(\vec{s}) + T_{B}(\vec{s} -
\vec{b}) \right]\ , \label{8}
\end{equation}

where $T_{AB}(b) = \int d^{2}\vec{s}\; T_{A}(\vec{s})
T_{B}(\vec{s} - \vec{b})$, $T_{A}(\vec{s}) = \int dz
\rho_{A}(\vec{s},z)$ is the nuclear density profile function,
$\rho_{A}(\vec{s},z)$ the nuclear matter density distribution and
$b$ the impact parameter. How to obtain $\epsilon_{0}$ as a
function of $b$ will be presented further.

To estimate ${\cal N}_{h.g.}(\epsilon_{0})$ we follow the
description presented in \cite{prtu1}. We shall focus on the plane
$z=0$ ($z$ is a collision axis) and put $J/\Psi$ longitudinal
momentum equal to zero. Now the $p_{T}$-dependent $J/\Psi$
survival factor ${\cal N}_{h.g.}(p_{T})$ is given by (for details
see \cite{prtu1})

\begin{equation}
{\cal N}_{h.g.}(p_{T})= \int
d^{2}\vec{s} f_{0}(s,p_{T}) \exp \left\{ -\int_{t_{0}}^{t_{f}} dt
\sum_{i=1}^{l} \int { {d^{3}\vec{q}} \over {(2\pi)^{3}} }
f_{i}(\vec{q},t) \sigma_{i} v_{rel,i} { {p_{\nu}q_{i}^{\nu}} \over
{EE^{\prime}_{i}} } \right\}\ , \label{9}
\end{equation}

where the sum in the power is over all taken species of scatters
(hadrons), $p^{\nu}=(E,\vec{p}_{T})$ and
$q_{i}^{\nu}=(E^{\prime}_{i},\vec{q})$ are four momenta of
$J/\Psi$ and hadron specie $i$ respectively, $\vec{v}=
\vec{p}_{T}/E$ is the velocity of the former, $\sigma_{i}$ states
for the absorption cross-section of $J/\Psi-h_{i}$ scattering and
$v_{rel,i}$ is the relative velocity of $h_{i}$ hadron with
respect to $J/\Psi$. When $M$ denotes $J/\Psi$ mass, $M= 3097$
MeV, $v_{rel,i}$ reads

\begin{equation}
v_{rel,i}=\left( 1- { {m_{i}^{2}M^{2}} \over
{(p_{\nu}q_{i}^{\nu})^{2}} } \right)^{{1 \over 2}} \ . \label{10}
\end{equation}

The upper limit of the time integral in (\ref{9}) , $t_{f}$, is
equal to $t_{f.o.}$ or to $t_{esc}$ -- the moment of leaving by a
given $J/\Psi$ of the hadron medium, if the final-size effects are
considered and $t_{esc} < t_{f.o.}$. For $\sigma_{i}$ we have
assumed that it equals zero for $(p^{\nu}+q_{i}^{\nu})^{2} <
(2m_{D} + m_{X})^{2}$ and is constant elsewhere ($m_{D}$ is a
charm meson mass, $m_{D}= 1867$ MeV). For hadron specie $i$ we
have usual Bose-Einstein or Fermi-Dirac distribution (we neglect
any possible spatial dependence here)

\begin{equation}
f_{i}(\vec{q},t)=f_{i}(q,t)={ {2s_{i}+1} \over {
\exp \left\{ { { E^{\prime}_{i}-\mu_{i}(t)} \over {T(t)} }
\right\} + g_{i} } }\ . \label{11}
\end{equation}

In the following, we shall consider only $J/\Psi$ initial
distribution $f_{0}(s,p_{T})$ that factorizes into
$f_{0}(s)g(p_{T})$ and the momentum distribution $g(p_{T})$ will
be given by (\ref{4}) . We assume at the first step that the
transverse size of the hadron medium is much greater than
$t_{f.o.}$ and also much greater than the size of the area where
$f_{0}(s)$ is non-zero. Additionally we assume that $f_{0}(s)$ is
uniform and normalized to unity. Note that the first assumption
overestimates the suppression but the second, in the presence of
the first, has no any calculable effect here. As a result, ${\cal
N}_{h.g.}(p_{T})$ simplifies to

\begin{equation}
{\cal N}_{h.g.}(p_{T}) = g(p_{T},\epsilon_{0}) \cdot \exp \left\{
-\int_{t_{0}}^{t_{f.o.}} dt \sum_{i=1}^{l} \int { {d^{3}\vec{q}}
\over {(2\pi)^{3}} } f_{i}(\vec{q},t) \sigma_{i} v_{rel,i} {
{p_{\nu}q_{i}^{\nu}} \over {EE^{\prime}_{i}} } \right\}\ ,
\label{12}
\end{equation}

To obtain ${\cal N}_{h.g.}(\epsilon_{0})$ one needs only to
integrate (\ref{12})  over $p_{T}$:

\begin{equation}
{\cal N}_{h.g.}(\epsilon_{0}) = \int dp_{T}\;
g(p_{T},\epsilon_{0}) \cdot \exp \left\{ -\int_{t_{0}}^{t_{f.o.}}
dt \sum_{i=1}^{l} \int { {d^{3}\vec{q}} \over {(2\pi)^{3}} }
f_{i}(\vec{q},t) \sigma_{i} v_{rel,i} { {p_{\nu}q_{i}^{\nu}} \over
{EE^{\prime}_{i}} } \right\}\ . \label{13}
\end{equation}

Now we would like to take the final-size effects and the
transverse expansion into account in our model. To do this
directly, we would have to come back to the formula given by
(\ref{9}) and integrate it, instead of (\ref{12}) , over $p_{T}$.
But this would involve a five-dimensional integral (the
three-dimensional integral over $\vec{q}$ simplifies to the
one-dimensional one, in fact) instead of the three-dimensional
integral of (\ref{13}). Therefore, we need to simplify in some way
the direct method just mentioned above. We shall define an average
time of leaving the hadron medium by $J/\Psi$'s with the velocity
$v$ produced in an A-B collision at impact parameter $b$, $\langle
t_{esc}\rangle(b,v)$. Then, we will put this quantity, instead of
$t_{f.o.}$, as the upper limit of the integral over $t$ in
(\ref{13}).

 %% 2
 \begin{figure}
\begin{center}{
{\epsfig{file=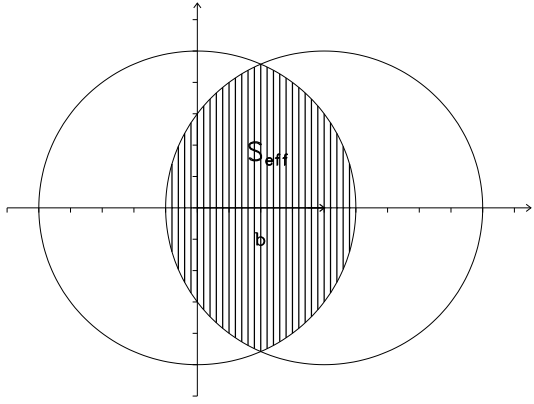,width=7cm}}
}\end{center}
\caption{View of a Pb-Pb collision at impact parameter $b$ in the
transverse plane ($z=0$). The region where the nuclei overlap has
been hatched and its area equals $S_{eff}$.}
\label{Fig.2.}%2
\end{figure}

Let us consider an A-B collision at impact parameter $b$. Since we
will compare final results with the latest data of NA50 which are
for Pb-Pb collisions \cite{Abr}, we focus on the case of A=B here.
So, for the collision at impact parameter $b$ we have the
situation in the plane $z=0$ as presented in Fig.~\ref{Fig.2.},
where $S_{eff}$ means the area of the overlap of the colliding
nuclei. We shall assume here, that the hadron medium, which
appears in the space between the nuclei after they crossed each
other also has the shape of $S_{eff}$ at $t_{0}$ in the plane
$z=0$. And additionally, the transverse expansion will start in
the form of the rarefaction wave moving inward $S_{eff}$ at
$t_{0}$. Then, for a $J/\Psi$ which is at $\vec{r} \in S_{eff}$ at
the moment $t_{0}$ and has the velocity $\vec{v}$ we denote by
$t_{esc}$ the moment of crossing the border of the hadron gas. It
means that $t_{esc}$ is a solution of the equation $\mid \vec{d} +
\vec{v} (t-t_{0}) \mid = R_{A} - c_{s} (t-t_{0})$, where $R_{A} =
r_{0} \cdot A^{{1 \over 3}}$ ($r_{0}=1.2 fm$) is the nucleus
radius and $\vec{d} = \vec{r} - \vec{b}$ for the angel between
$\vec{r}$ and $\vec{v}$ such that the $J/\Psi$ will cross this
part of the edge of the area of the hadron gas which was created
by the projectile and $\vec{d} = \vec{r}$ in the opposite. Having
obtain $t_{esc}$, we average it over the angel between $\vec{r}$
and $\vec{v}$, i.e. we integrate $t_{esc}$ over this angel and
divide by $2\pi$. Then we average the result over $S_{eff}$ with
the weight given by

\begin{equation}
p_{J/\Psi}(\vec{r}) = {
{T_{A}(\vec{r})T_{B}(\vec{r} - \vec{b})} \over {T_{AB}(b)} }
\label{14}
\end{equation}

and we obtain $\langle t_{esc}\rangle(b,v)$. So, the final
expression for ${\cal N}_{h.g.}(\epsilon_{0})$ when the transverse
expansion is taken into account reads

\begin{equation}
{\cal N}_{h.g.}(\epsilon_{0}) = \int dp_{T}\;
g(p_{T},\epsilon_{0}) \cdot \exp \left\{ -\int_{t_{0}}^{\langle
t_{esc}\rangle} dt \sum_{i=1}^{l} \int { {d^{3}\vec{q}} \over
{(2\pi)^{3}} } f_{i}(\vec{q},t) \sigma_{i} v_{rel,i} {
{p_{\nu}q_{i}^{\nu}} \over {EE^{\prime}_{i}} } \right\}\ .
\label{15}
\end{equation}

\section { The energy density in the CRR }

To compare our theoretical estimations for $J/\Psi$ survival
factor with the experimental data \cite{Abr} we need to modify the
latest so as they become $\epsilon_{0}$-dependent instead of
$E_{T}$-dependent ($E_{T}$ is the neutral transverse energy). To
do this we will use the well-known Bjorken formula

\begin{equation}
\epsilon_{0} = { {3 \cdot E_{T}} \over {
\Delta\eta S_{eff} t_{0} } }\ , \label{16}
\end{equation}

where $\Delta\eta$ is the pseudo-rapidity range. Using values of
impact parameter $b$ given in \cite{Abr} we can calculate
$S_{eff}$ for each measured $E_{T}$ bin.

In further considerations we will need the formula for the number
of participating nucleons as a function of impact parameter $b$:

\begin{equation}
N_{part}(b) = \int_{S_{eff}} d^{2}\vec{s} \left\{
T_{A}(\vec{s}) + T_{B}(\vec{s} - \vec{b}) \right\}\ . \label{17}
\end{equation}

We found out that the ratio $E_{T}$/$N_{part}$ is almost constant
and for NA50 data \cite{Abr} varies from 0.28 GeV to 0.31 GeV with
the average value equal roughly to 0.3 GeV and the standard
deviation equal to 0.0094 GeV. Therefore we assume that for Pb-Pb
collisions of NA50 the following approximation is valid:

\begin{equation}
E_{T} \cong 0.3 \cdot N_{part}\ . \label{18}
\end{equation}

Having put (\ref{18})  into (\ref{16})  we obtain $\epsilon_{0}$
as a function of $b$

\begin{equation}
\epsilon_{0}(b) = 0.75 \cdot {
{N_{part}(b)} \over {S_{eff}(b)} }\ , \label{19}
\end{equation}

\begin{figure}
\begin{center}{
{\epsfig{file=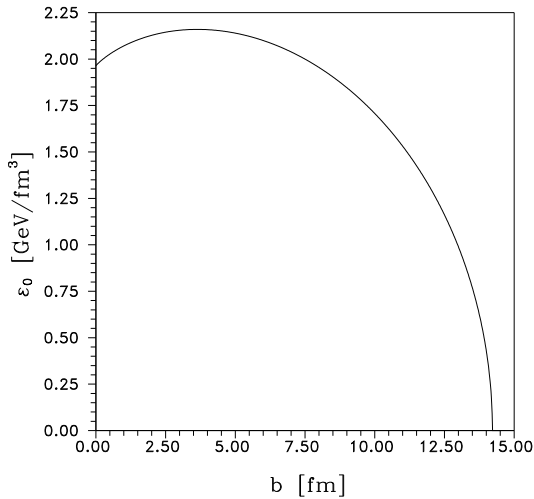,width=7cm}}
}\end{center}
\caption{The initial energy density $\epsilon_0$ in the CRR for
Pb-Pb collisions as the function of impact parameter $b$ and for
the NA50 data of 1996 run \protect\cite{Abr}}
\label{Fig.3.}%3
\end{figure}
where we have also used the value $\Delta\eta=1.2$ of NA50
\cite{Abr}. The above function is depicted in Fig.~\ref{Fig.3.}.
The behaviour in low $b$ is the most interesting feature of
$\epsilon_{0}(b)$. We can see that for $b \leq 7.9$ there are two
different values $b_{1}$ and $b_{2}$ such that
$\epsilon_{0}(b_{1}) = \epsilon_{0}(b_{2})$. Of course, this is
the result of direct application of (\ref{18})  which is some
approximation in fact. But nevertheless this can suggest that
there is a wide range of impact parameter $b$ for which the
resulting $\epsilon_{0}$ is almost constant. Using the dependence
between $E_{T}$ and $b$ obtained by NA50 \cite{Abr}, we can state
the similar for $E_{T}$, i.e. for $E_{T} \geq 40-50 GeV$
$\epsilon_{0}$ changes weakly with $E_{T}$. Also the same is true
for $J/\Psi$ suppression what could suggest that the quantity of
$\epsilon_{0}$ reached is the main reason for the suppression.

\section { Results }

To evaluate formulae (\ref{13})  and (\ref{15})  we have to know
$T(t)$, $\mu_{B}(t)$ and $\mu_{S}(t)$ and how to obtain these
functions was explained in Sect.~\ref{hadgas}. But to follow all
that procedure we need initial values $s_{0}$ and $n_{B}^{0}$. To
estimate initial baryon number density $n_{B}^{0}$ we can use
experimental results for S-S \cite{Stro} or Au-Au \cite{Stach,Ahl}
collisions. In the first approximation we can assume that the
baryon multiplicity per unit rapidity in the CRR is proportional
to the number of participating nucleons. For a sulphur-sulphur
collision we have $dN_{B}/dy \cong 6$ \cite{Stro} and 64
participating nucleons. For the central collision of lead nuclei
we can estimate the number of participating nucleons at $2A =
416$, so we have $dN_{B}/dy \cong 39$. Having taken the initial
volume in the CRR equal to $\pi R_{A}^{2} \cdot 1$ fm, we arrive
at $n_{B}^{0} \cong 0.25 \; fm^{-3}$. This is some underestimation
because the S-S collisions were at a beam energy of 200
GeV/nucleon, but Pb-Pb at 158 GeV/nucleon. From the Au-Au data
extrapolation one can estimate $n_{B}^{0} \cong 0.65 \; fm^{-3}$
\cite{Stach}. These values are for central collisions, and for the
higher impact parameter (a more peripheral collision) the initial
baryon number density should be much lower. In fact, if we apply
the above-mentioned assumption, the initial baryon number density
for a given collision at the impact parameter $b$ will be
proportional to the number of participating nucleons divided by
$S_{eff}$. Therefore, $n_{B}^{0}(b)$ will have exactly the same
shape as $\epsilon_{0}(b)$ presented in Fig.~\ref{Fig.3.}. As a
result, only for the most peripheral collisions $n_{B}^{0}$ will
be substantially below the value for the central one. So, to
simplify numerical calculations we will keep $n_{B}^{0}$ constant
over the all range of $b$ and additionally, to check the possible
dependence on $n_{B}^{0}$, we will do our estimations for
$n_{B}^{0}$ substantially lower, i.e. $n_{B}^{0} = 0.05 fm^{-3}$.
According to our approximation of $n_{B}^{0}(b)$, it would
%%%%%%%%%%%%%%%%%%%%%%%%%%%%%%%%%%%%%%%%%%%%%%%%%%%%%%%%%%%%%%%%%%%%%%%%%%%%%%
%%%%%%%%%%%uzupe³nienie
%%%%%%%%%%%%%%%%%%%%%%%%%%%%%%%%%%%%%%%%
correspond to the most peripheral Pb-Pb collisions, $b \geq 14
fm$.

Now, to find $s_{0}$, first we have to solve (\ref{2a} - \ref{2c})
with respect to $T$, $\mu_{S}$ and $\mu_{B}$, where we put
$\epsilon = \epsilon_{0}$, $n_{B} = n_{B}^{0}$ and $n_{S}=0$.
Then, having put $T$, $\mu_{S}$ and $\mu_{B}$ into (\ref{2d})  we
obtain $s_{0}$. Finally, expressing left sides of
(\ref{2b},\ref{2d}) by (\ref{3})  and after then solving (\ref{2b}
- \ref{2d})  numerically we can obtain $T$, $\mu_{S}$ and
$\mu_{B}$ as functions of time. In fact, evaluating formulae
(\ref{13})  and (\ref{15})  we do the following: first, we
calculate $T=T(t)$ which turns out to be very well approximated by
the expression

\begin{equation}
T(t) \cong T_{0} \cdot t^{-a} \label{20}
\end{equation}

and then we put this approximation into (\ref{13})  and (\ref{15})
. And for $\mu_{S}(t)$ and $\mu_{B}(t)$ in $f_{i}(\vec{q},t)$ we
put solutions of (\ref{2b},\ref{2c})  where $n_{B}$ given by
(\ref{3}) and $n_{S}=0$ and $T$ is given by (\ref{20}). But the
exponent $a$ in (\ref{20})  has proven not to be unique for the
whole range of $T_{0}$ considered here. One gets different values
of the initial energy density $\epsilon_0$ for different values of
the impact parameter $b$  and for different geometry of the
collision process. So $b$ dependent $a$ gives also $b$ dependent
freeze-out time $t_{f.o}$. The formula (\ref{19})  is calculated
for different values of $b$ and Eqs. (\ref{2a} - \ref{2c})  are
solved. We have evaluated the suppression factor up to
$\epsilon_{0} = 3.5 GeV/fm^{3}$. This gives e.g. the maximal
possible $T_{0}$, $T_{0,max}$, equal to $219.3 MeV$ ( for
$n_{B}^{0} = 0.65 fm^{-3}$), $224 MeV$ (for $n_{B}^{0} = 0.25
fm^{-3}$) or $224.8 MeV$ (for $n_{B}^{0} = 0.05 fm^{-3}$).

This procedure allows to evaluate $J/\Psi$ survival factor given
by (\ref{13}) . Because of the lack of data, we shall assume only
two types of the cross-section, the first, $\sigma_{b}$, for
$J/\Psi$-baryon scattering and the second, $\sigma_{m}$, for
$J/\Psi$-meson scattering. For $\sigma_{b}$ we put $\sigma_{b} =
\sigma_{J/\psi N}$. As far as $\sigma_m$ is concerned, we assume
that this cross-section is $2/3$ of the corresponding cross
section for baryons, which is due to the quark counting. In the
following, we will use values of $J/\Psi-Nucleon$ absorption
cross-section $\sigma_{J/\psi N} \cong 3-5 mb$ obtained from p-A
data \cite{Bad,Gers,GavVo}. At the beginning, to illustrate how
the value of power $a$ influences $J/\Psi$ suppression we present
in Fig.~\ref{Fig.1.} two results: the first for $a={1 \over 3}$
(which is the exact value for a free massless gas) and the second
for $a={1 \over {5.6}}$ (which is the approximate value for the
hadron gas and $T_{0} \cong 200 MeV$). We can see that the
suppression improves more than twice for the highest
$\epsilon_{0}$ indeed.
%%45
\begin{figure}[htb]
\begin{minipage}[t]{75mm}
{\psfig{figure= 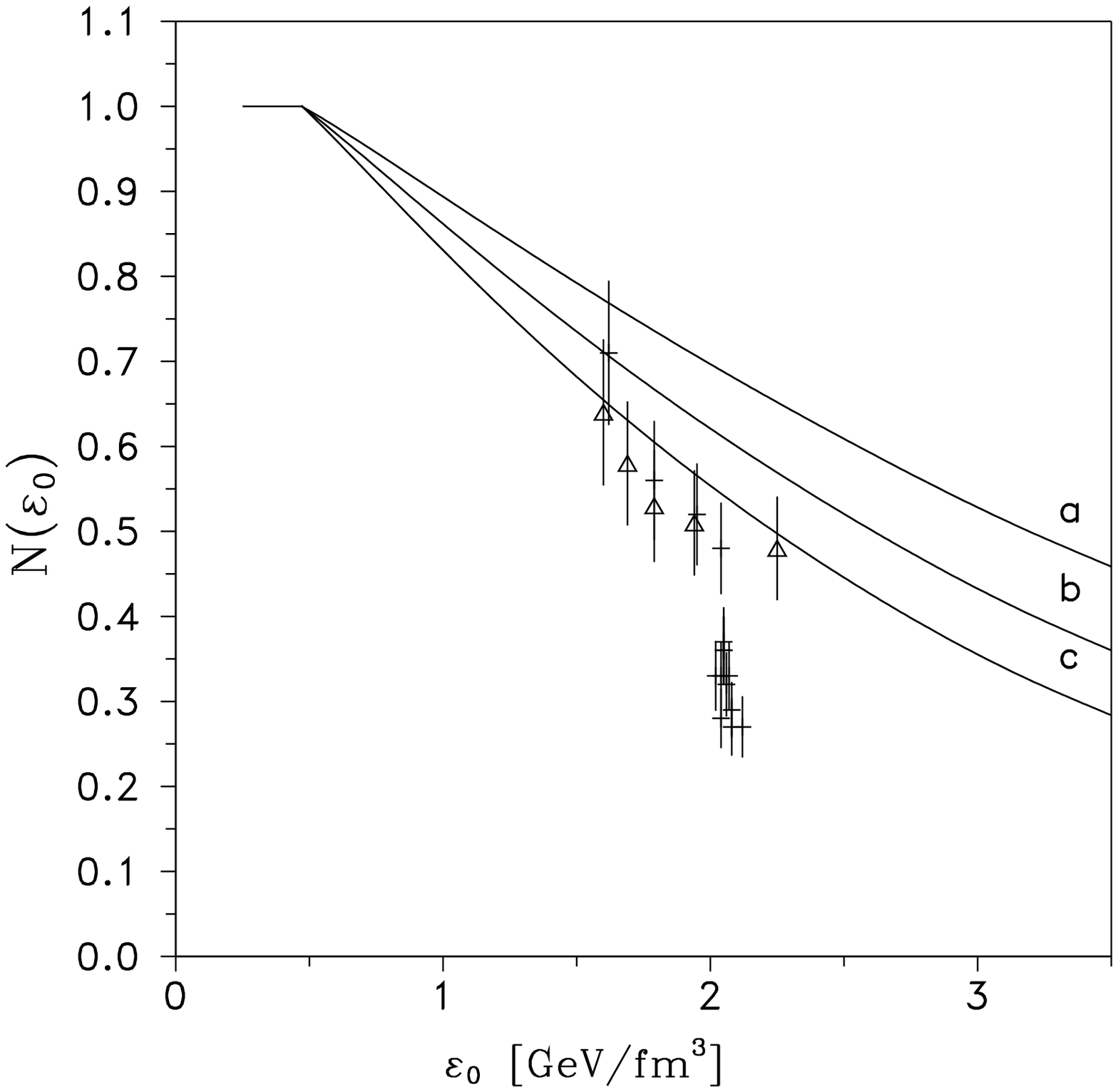,height=\textwidth,width=\textwidth}}
\caption{$J/\Psi$ suppression in the longitudinally expanding
hadron gas with the "infinite" transverse size and for
$n_{B}^{0}=0.25\;fm^{-3}$ and $T_{f.o.}=140\;MeV$: a)
$\sigma_{b}=3\;mb$, $\sigma_{m}=2\;mb$; b) $\sigma_{b}= 4\;mb$,
$\sigma_{m}=2.66\;mb$; c) $\sigma_{b}=5\;mb$,
$\sigma_{m}=3.33\;mb$. Triangles and crosses represent the S-U
data of NA38 and the Pb-Pb data of NA50 Collaborations
\protect\cite{Abr}, respectively.} \label{Fig.4.}
\end{minipage}
\hspace{\fill}
\begin{minipage}[t]{75mm}
{\psfig{figure= 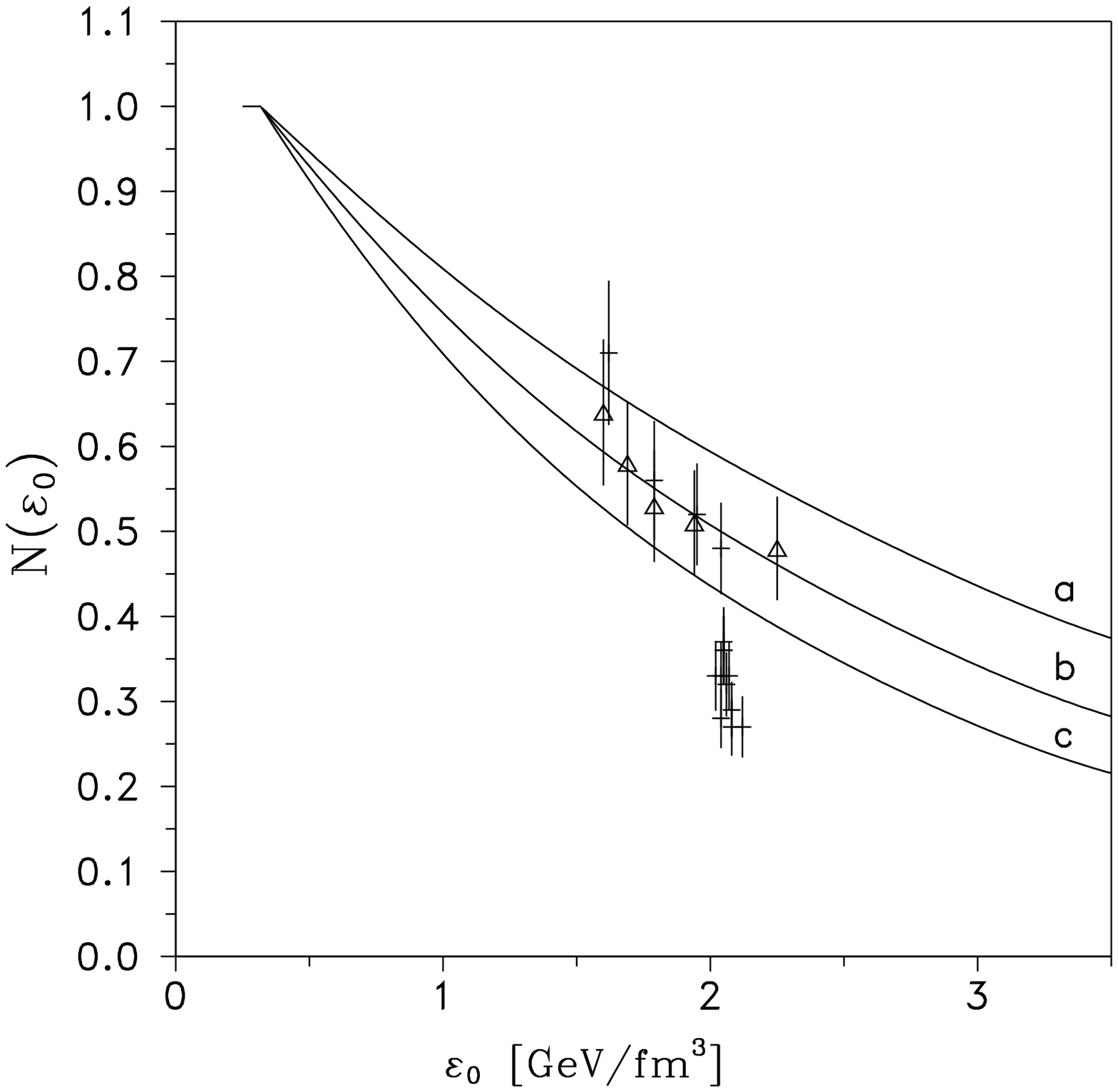 ,height=\textwidth,width=\textwidth}}
\caption{Same as Fig.~\ref{Fig.4.} but for$T_{f.o.}=100\;MeV$.}
\label{Fig.5.}
\end{minipage}
\end{figure}

To make our investigations more realistic we have to take into
account that only about $60 \%$ of $J/\Psi$ measured are directly
produced during collision. The rest is the result of $\chi$ ($\sim
30 \%$) and $\psi'$ ($\sim 10 \%$) decay \cite{Satz}. Therefore
the realistic $J/\Psi$ survival factor should read

\begin{equation}
{\cal N}(\epsilon_{0})=0.6{\cal
N}_{J/\psi}(\epsilon_{0})+0.3{\cal N}_{\chi}(\epsilon_{0})+
0.1{\cal N}_{\psi'}(\epsilon_{0})\;, \label{21}
\end{equation}

where ${\cal N}_{J/\psi}(\epsilon_{0})$, ${\cal
N}_{\chi}(\epsilon_{0})$ and ${\cal N}_{\psi'}(\epsilon_{0})$ are
given also by formulae (\ref{4} -\ref{15})  but with $\langle
p_{T}^{2}\rangle_{J/\Psi}^{AB}(\epsilon)=\langle
p_{T}^{2}\rangle_{J/\Psi}^{AB}(\epsilon), \langle
p_{T}^{2}\rangle_{\chi}^{AB}(\epsilon), \langle
p_{T}^{2}\rangle_{\psi'}^{AB}(\epsilon)$, $K_{J/\psi}=K_{J/\psi},
K_{\chi}, K_{\psi'}$, $\sigma_{J/\psi N}=\sigma_{J/\psi N},
\sigma_{\chi N}, \sigma_{\psi' N}$ and $M=M_{J/\psi}, M_{\chi},
M_{\psi'}$ respectively. The remaining problem is whether formula
(\ref{5})  is valid for $\chi$ and $\psi'$. There are data for
$\langle p_{T}^{2}\rangle_{\psi'}^{PbPb}$ \cite{Abr98} and they
shows that $\langle p_{T}^{2}\rangle_{\psi'}^{PbPb} \approx 1.4
\langle p_{T}^{2}\rangle_{J/\Psi}^{PbPb}$. So, we assume that the
above is also true for $\langle
p_{T}^{2}\rangle_{\psi'}^{AB}(\epsilon)$, i.e.

\begin{equation}
\langle p_{T}^{2}\rangle_{\psi'}^{AB}(\epsilon)=1.4 \langle
p_{T}^{2}\rangle_{J/\Psi}^{AB}(\epsilon) \label{22}
\end{equation}

with $\langle p_{T}^{2}\rangle_{J/\Psi}^{AB}(\epsilon)$ given by
(\ref{5}) . For $\chi$ we believe that the inequality

\begin{equation}
\langle p_{T}^{2}\rangle_{J/\Psi}^{AB}\;\; \leq \;\; \langle
p_{T}^{2}\rangle_{\chi}^{AB}\;\; \leq \;\; \langle
p_{T}^{2}\rangle_{\psi'}^{AB} \label{23}
\end{equation}

should be valid and therefore assume that (\ref{22})  is true also
in this case. Anyway, the exact form of $\langle
p_{T}^{2}\rangle_{\chi}^{AB}(\epsilon)$ or $\langle
p_{T}^{2}\rangle_{\psi'}^{AB}(\epsilon)$ is not very important
because we checked that the suppression depends on this form very
weakly. First, we put $K_{J/\psi} = 0$ and the resulting $J/\Psi$
survival factor (for direct $J/\Psi$'s) differs only a few percent
for the highest $\epsilon_{0}$ from the one calculated with
formula (\ref{5})  unchanged. Second, when we use expression
(\ref{5}) also for $\chi$ and $\psi'$, the evaluated suppression
factor is the same as that calculated with the use of (\ref{22}),
as far as plots are concerned.

\begin{figure}[htb]
\begin{minipage}[t]{75mm}
{\psfig{figure= 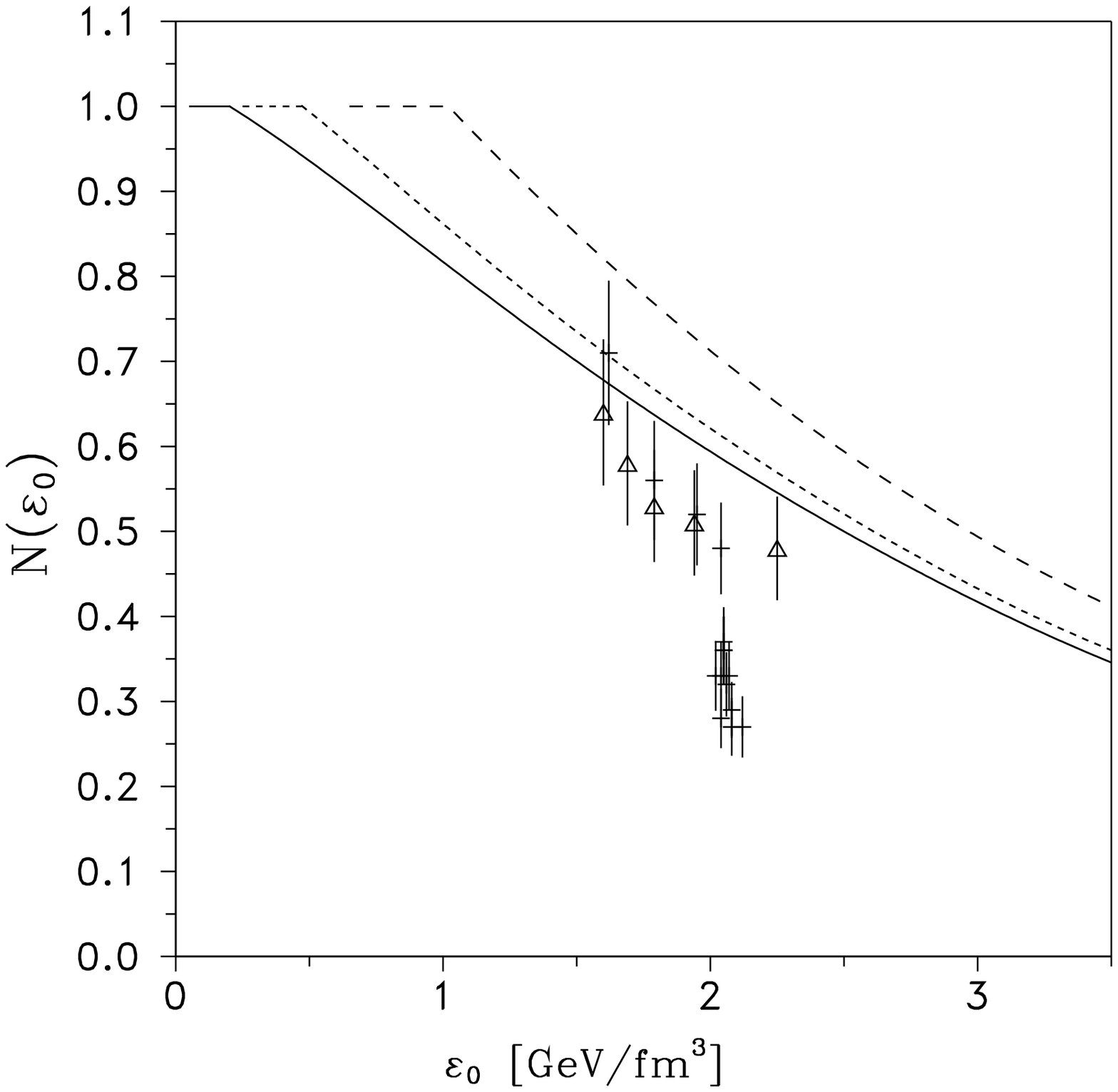,height=\textwidth,width=\textwidth}}
\caption{$J/\Psi$ suppression in the longitudinally expanding
hadron gas with the "infinite" transverse size and for
$\sigma_{b}=4\;mb$, $\sigma_{m}=2.66\;mb$ and $T_{f.o.}=140\;MeV$.
The curves correspond to $n_{B}^{0}=0.05\;fm^{-3}$ (solid),
$n_{B}^{0}=0.25\; fm^{-3}$ (short-dashed) and
$n_{B}^{0}=0.65\;fm^{-3}$ (dashed). Triangles and crosses
represent the S-U data of NA38 and the Pb-Pb data of NA50
Collaborations \protect\cite{Abr}, respectively.} \label{Fig.6.}
\end{minipage}
\hspace{\fill}
\begin{minipage}[t]{75mm}
{\psfig{figure= 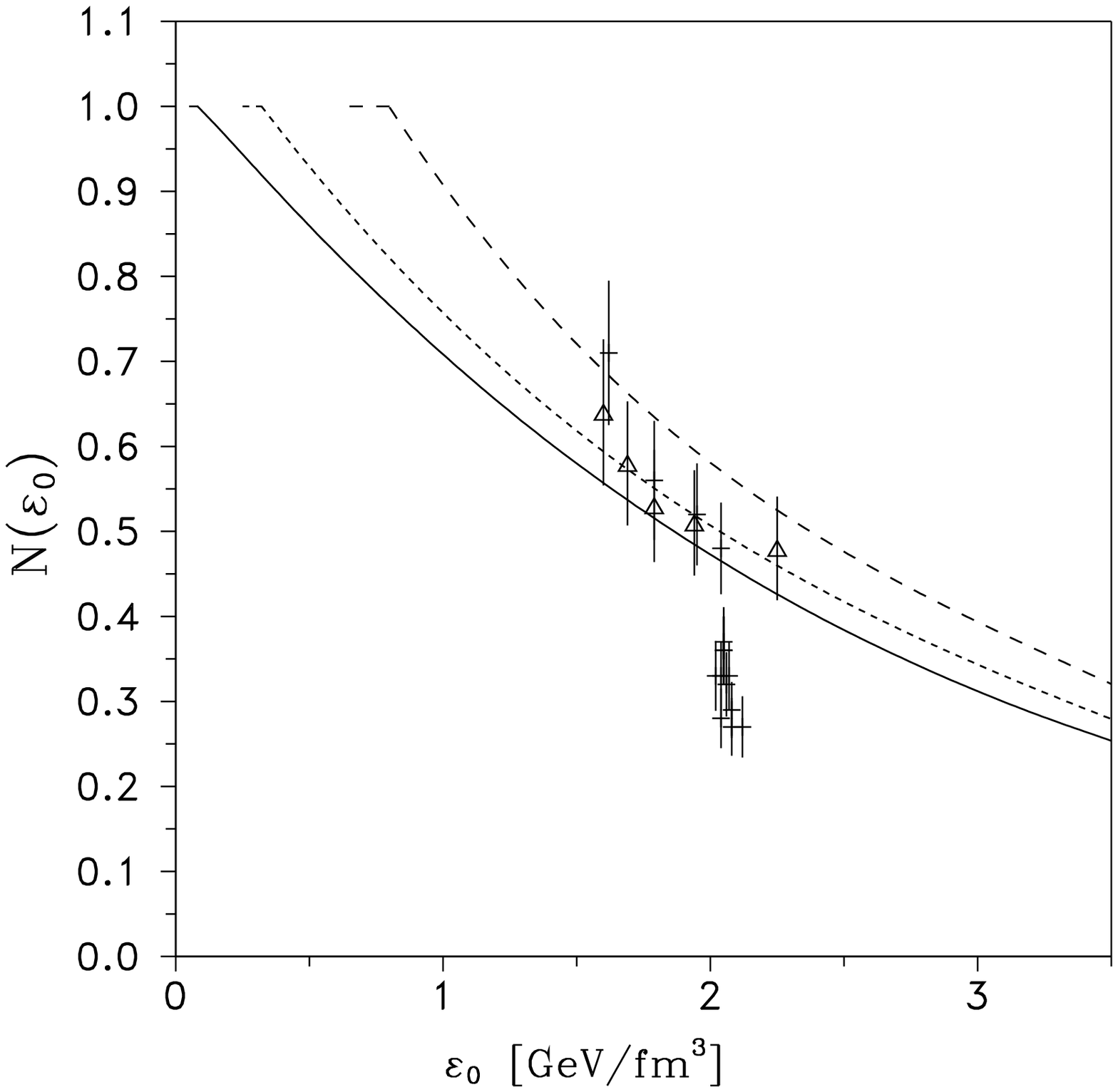 ,height=\textwidth,width=\textwidth}}
\caption{Same as Fig.~\ref{Fig.6.} but for $T_{f.o.}=100\;MeV$.}
\label{Fig.7.}
\end{minipage}
\end{figure}
To complete our estimations we need also values of cross-sections
for $\chi-baryon$ and $\psi'-baryon$ scatterings (we will still
hold that $\chi(\psi')-meson$ cross-section is ${2 \over 3}$ of
$\chi(\psi')-baryon$ cross-section). Since $J/\Psi$ is smaller
than $\chi$ or $\psi'$, $\chi-baryon$ and $\psi'-baryon$
cross-sections should be greater than $J/\Psi-baryon$ one. For
simplicity, we assume that all these cross-sections are equal.
This means that we {\it underestimate} $J/\Psi$ suppression, here.
The final results of calculations of (\ref{13})  are presented in
Figs.~\ref{Fig.4.}-\ref{Fig.7.} for various sets of parameters of
our model (which are $T_{f.o.}, n_{B}^{0}, \sigma_{b}$). We
performed these calculations for two values of $T_{f.o.}=100, 140
MeV$ which agree fairly well with values deduced from hadron
yields \cite{Stach}. For comparison, also the experimental data
are shown in Figs.~\ref{Fig.4.}-\ref{Fig.7.}. The experimental
survival factor is defined as

\begin{equation}
{\cal N}_{exp}={ { {B_{\mu\mu}\sigma_{J/\psi}^{AB}} \over
{\sigma_{DY}^{AB}} } \over { {B_{\mu\mu}\sigma_{J/\psi}^{pp}}
\over {\sigma_{DY}^{pp}}  } }\;\;, \label{24}
\end{equation}

where ${B_{\mu\mu}\sigma_{J/\psi}^{AB(pp)}} \over
{\sigma_{DY}^{AB(pp)}}$ is the ratio of the $J/\Psi$ to the
Drell-Yan production cross-section in A-B(p-p) interactions times
the branching ratio of the $J/\Psi$ into a muon pair. The values
of the ratio for p-p, S-U and Pb-Pb are taken from \cite{Abr}.
Note that since the equality $\sigma_{DY}^{AB}=\sigma_{DY}^{pp}
\cdot AB$ has been confirmed experimentally up to now
\cite{Abr98}, formula (\ref{24})  reduces to

\begin{equation}
{\cal N}_{exp}= { {\sigma_{J/\psi}^{AB}} \over {AB
\sigma_{DY}^{pp}} }\;\;, \label{25}
\end{equation}

which is also given as the experimental survival factor, for
instance, in \cite{Lou}.

\begin{figure}[htb]
\begin{minipage}[t]{75mm}
{\psfig{figure= 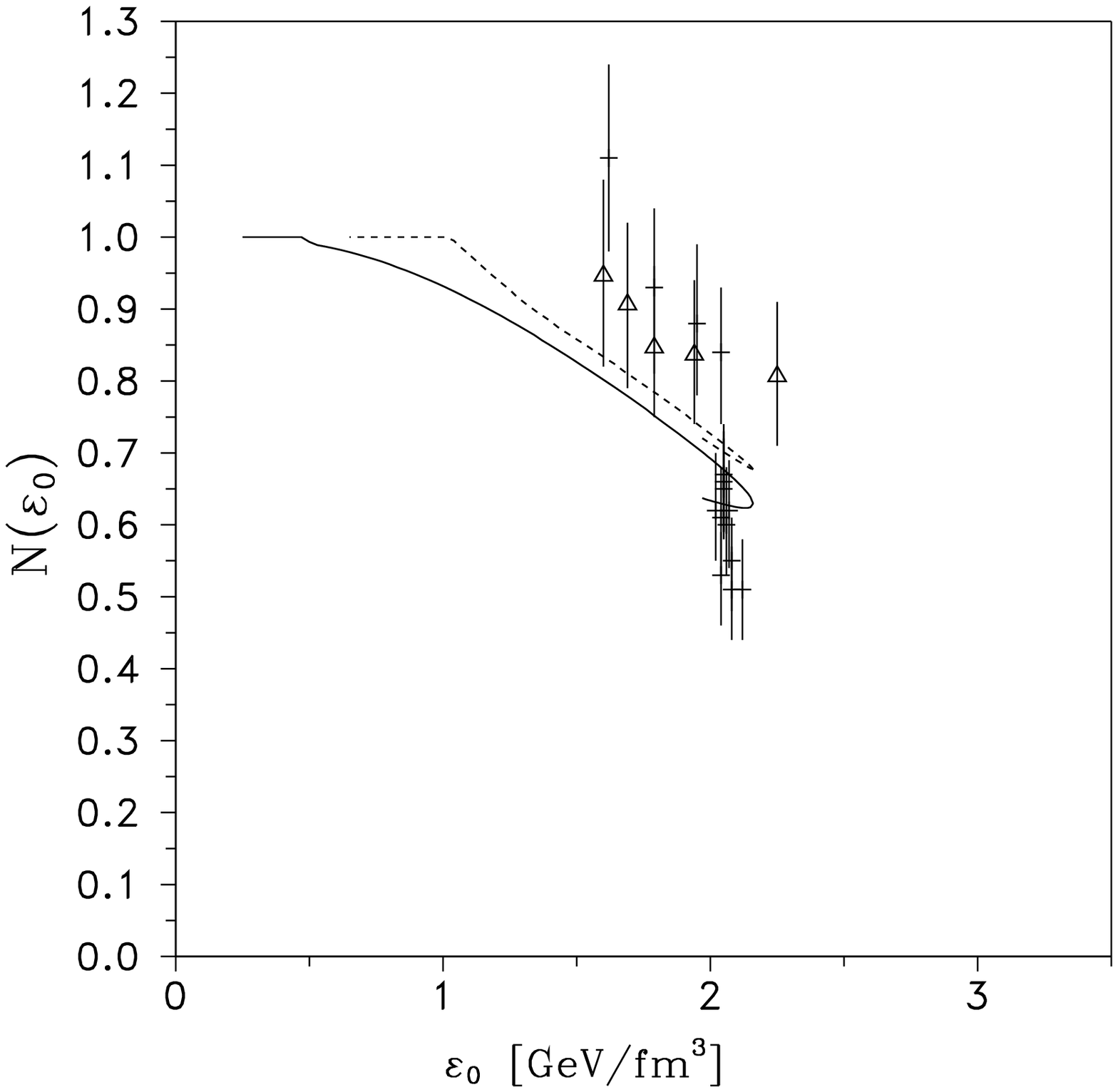,height=\textwidth,width=\textwidth}}
 \caption{$J/\Psi$ suppression in the longitudinally
and transversely expanding hadron gas for the uniform nuclear
matter density distribution and $\sigma_{b}=4\;mb$,
$\sigma_{m}=2.66\;mb$ and $T_{f.o.}=140\;MeV$. The curves
correspond to $n_{B}^{0}=0.25\;fm^{-3}$, $c_{s}=0.45$ (solid) and
$n_{B}^{0}=0.65\;fm^{-3}$, $c_{s}=0.46$ (dashed). Triangles and
crosses represent the S-U data of NA38 and the Pb-Pb data of NA50
Collaborations \protect\cite{Abr} respectively, but the data are
"cleaned out" from the contribution of $J/\Psi$ scattering in the
nuclear matter, in accordance with (~\ref{26}~)}. \label{Fig.8.}
\end{minipage}
\hspace{\fill}
\begin{minipage}[t]{75mm}
{\psfig{figure= 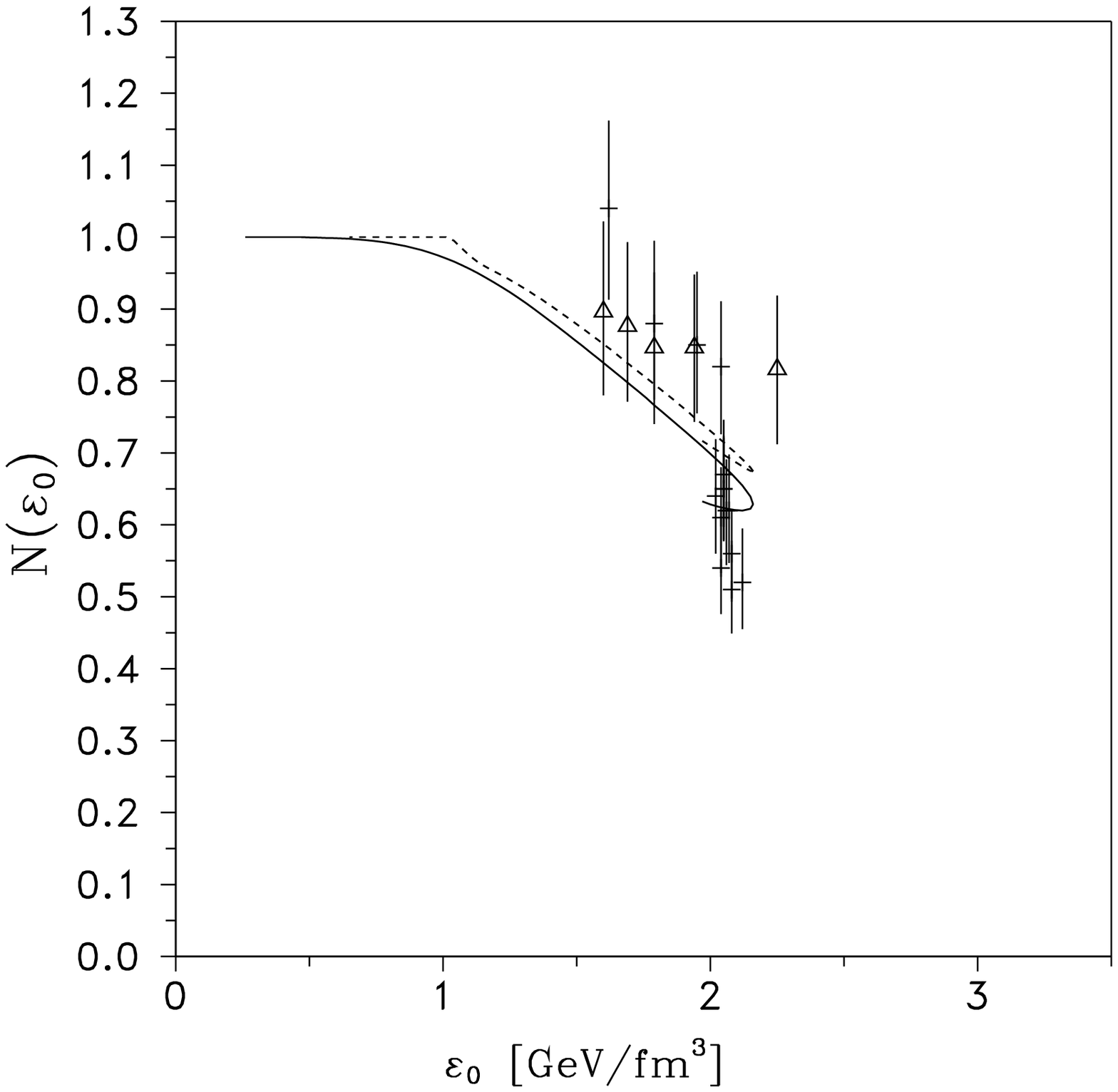 ,height=\textwidth,width=\textwidth}}
 \caption{$J/\Psi$ suppression in the longitudinally and
transversely expanding hadron gas for the Woods-Saxon nuclear
matter density distribution and $\sigma_{b}=4\;mb$,
$\sigma_{m}=2.66\;mb$ and $T_{f.o.}=140\;MeV$. The curves
correspond to $n_{B}^{0}=0.25\;fm^{-3}$, $c_{s}=0.45$ (solid) and
$n_{B}^{0}=0.65\;fm^{-3}$, $c_{s}=0.46$ (dashed). Triangles and
crosses represent the S-U data of NA38 and the Pb-Pb data of NA50
Collaborations \protect\cite{Abr} respectively, but the data are
"cleaned out" from the contribution of $J/\Psi$ scattering in the
nuclear matter, in accordance with (~\ref{26}~).} \label{Fig.9.}
\end{minipage}
\end{figure}

Coming back to examination of our first results presented in
Figs.~\ref{Fig.4.}-\ref{Fig.7.}, we can see that the most of Pb-Pb
data are below the region of suppression obtained for chosen
values of parameters in our model. Generally, the theoretical
curves decrease slower with increasing $\epsilon_{0}$, whereas the
data show rather abrupt fall just above $\epsilon_{0}=2
GeV/fm^{3}$. Note that the dependence on the initial baryon number
density is substantial but for higher values of $n_{B}^{0}$,
rather. The lower the initial baryon number density, the deeper
the suppression. There are two reasons for such a behaviour: the
first, for the higher baryon number density, there are less
non-strange heavier mesons $\rho$, $\omega$ in the hadron gas of
the same $\epsilon_{0}$, but these particles create the most
weighty fraction of scatters, for which reaction (\ref{1}) have no
threshold at all; the second, the freeze-out time $t_{f.o.}$
decreases with increasing $n_{B}^{0}$ for a given $\epsilon_{0}$
in our model. For instance, for $\epsilon_{0}=3.5 GeV/fm^{3}$ and
$T_{f.o.}=140 MeV$ we have $a=0.172,\;0.175,\;0.183$ and $t_{f.o.}
= 15.7,\;14.7,\;11.6 fm$ for $n_{B}^{0}= 0.05,\;0.25,\;0.65
fm^{-3}$, respectively. We can see also that the value
$\sigma_{b}=3 mb$ is too small to obtain results comparable with
the data, so we will leave aside this value in further
investigations.

Now we will include the finite-size effects into our model, i.e.
we will take into account that the realistic hadron gas has a
finite transverse size. This will be done in form of the
rarefaction wave moving inward $S_{eff}$ with the sound velocity
$c_{s}$. How to obtain this velocity has been mentioned in Sec.2
(see also \cite{prtu3}). With the finite-size effects included,
the final expression for $J/\Psi$ survival factor ${\cal
N}_{h.g.}(\epsilon_{0})$ will be given by (\ref{15}) . To make our
investigations much more realistic we will also include the
possible $J/\Psi$ disintegration in nuclear matter, which should
increase $J/\Psi$ suppression by about $10 \%$ \cite{Gers}. But to
draw also S-U data in figures, instead of multiplying ${\cal
N}_{h.g.}$ by ${\cal N}_{n.m.}$ given by (\ref{7}), we divide
${\cal N}_{exp}$ by appropriate ${\cal N}_{n.m.}$, i.e. we define
"the experimental $J/\Psi$ hadron gas survival factor" as

\begin{figure}
\begin{center}{
{\epsfig{file=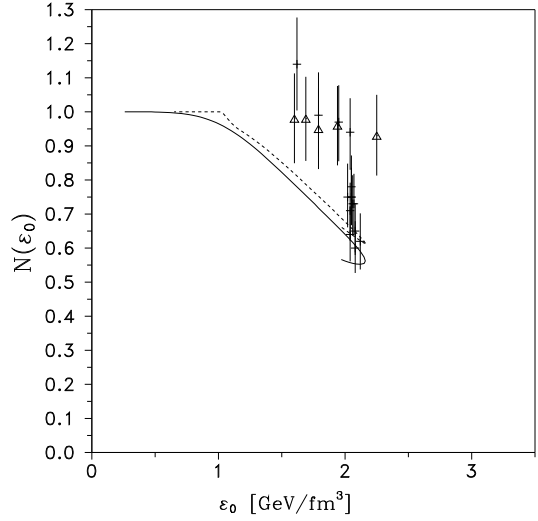,width=7cm}}
}\end{center}
\caption{Same as Fig.~\protect\ref{Fig.9.} but for
$\sigma_{b}=5\;mb$, $\sigma_{m}=3.33\;mb$.} \label{Fig.10.}
\end{figure}

\begin{equation}
\tilde{{\cal N}}_{exp}= \exp \left\{ \sigma_{J/\psi N} \rho_{0} L
\right\} \cdot {\cal N}_{exp}\;. \label{26}
\end{equation}

and values of this factor are drawn in
Figs.~\ref{Fig.8.}-\ref{Fig.10.} as the experimental data.

At the beginning we will consider a uniform nuclear matter density

\begin{equation}
\rho_{A}(\vec{s},z)=\rho_{A}(\vec{r})=\cases{ \rho_{0}= \left( {
{4\pi} \over 3} r_{0}^{3} \right)^{-1},\;\;\mid \vec{r} \mid \leq
R_{A} \cr \;\;\;\;0\;\;\; \;\;\;\;\;\;\;\;\;\;\;\;\;,\;\;\mid
\vec{r} \mid > R_{A} }\;\;\;\;\;. \label{27}
\end{equation}

The results of numerical estimations of (\ref{15})  and (\ref{26})
are depicted in Fig.~\ref{Fig.8.} for two values of the initial
baryon number density, $n_{B}^{0}=0.25,\;0.65\; fm^{-3}$. The
curve for $n_{B}^{0}=0.05\;fm^{-3}$ almost covers the curve for
$n_{B}^{0}=0.25\;fm^{-3}$ and the maximal difference between these
curves do not exceed $1.9 \%$ (which is the difference for the
highest possible $\epsilon_{0}$, i.e. $\epsilon_{0} \approx 2
GeV/fm^{-3}$), so for clearness of the figure we do not draw it.
The two values of the speed of sound are the maximal values of
this quantity possible in the range
$[T_{f.o.}=140\;MeV,T_{0,max}]$ for the above-mentioned two cases
of $n_{B}^{0}$. In fact, we have checked that the results almost
do not depend on $c_{s}$ (allowed in the range) and the difference
(seen only for the quantity $\epsilon_{0}$) between $J/\Psi$
survival factors for the maximal and the minimal values of $c_{s}$
in the range $[T_{f.o.}=140\;MeV,T_{0,max}]$ are less than $0.6
\%$. Note that the theoretical curves in Figs.~\ref{Fig.8.} (the
same will happen in Figs.~\ref{Fig.9.}-\ref{Fig.10.}) are
two-valued around $\epsilon_{0}= 2GeV/fm^{-3}$. This is the result
of our approximation of $\epsilon_{0}(b)$ given by (\ref{19}).
This expression allows for two different values of $b$, which give
the same $\epsilon_{0}$ in some range of the impact parameter (see
Sec.4).

It has turned out also that in the case of the transverse
expansion, the results almost do not depend on the $T_{f.o.}$ (for
$T_{f.o.} \in [100, 140] MeV$). And the maximal difference (seen
for the quantity $\epsilon_{0}$) between curves for $T_{f.o.} =100
MeV$ and $T_{f.o.}=140 MeV$ do not exceed $2.7 \%$
($n_{B}^{0}=0.05 fm^{-3}$), $2.6 \%$ ($n_{B}^{0}=0.25 fm^{-3}$).
This is because the freeze-out time resulting from the transverse
expansion, $t_{f.o.,trans} = R_{A}/c_{s}$ (if we assume a central
collision and $c_{s}$ constant), is of the order of the freeze-out
time resulting from the longitudinal expansion for $T_{f.o.}=140
MeV$. Namely, for Pb and $c_{s}=0.45$ we have $t_{f.o.,trans}
\cong 15.8 fm$ which is very similar to values of $t_{f.o.}$ for
$T_{f.o.}=140 MeV$ given earlier. For $T_{f.o.}=100 MeV$,
$t_{f.o.} = 111.0,\;101.0,\;72.5 fm$ for $n_{B}^{0}=
0.05,\;0.25,\;0.65 fm^{-3}$ respectively, so the hadron gas ceases
because of the transverse expansion much earlier.

We repeated our estimations of formula (\ref{15})  also for the
more realistic nuclear matter density distribution, namely for the
Woods-Saxon distribution with parameters taken from \cite{Jager}.
The results are presented in Figs.~\ref{Fig.9.}-\ref{Fig.10.}. We
can see that these curves fit the data a little bit better than
those obtained within the assumption of the uniform nuclear matter
density distribution (cf. Fig.~\ref{Fig.8.} and \ref{Fig.9.}).
Generally, taking into account also the transverse expansion
changes the final (theoretical) pattern of $J/\Psi$ suppression
qualitatively. First of all, the curves for the case including the
transverse expansion are concave (what is more clearly seen for
lower $n_{B}^{0}$) as the data suggest should be, in opposite to
the case with the longitudinal expansion only, where curves are
convex. But still, theoretical curves are not steep enough to
cover the data area completely. We would like to note at this
point that there is some ambiguity in calculation of $\langle
E_{T}\rangle$ and $b$ in NA50 experiment, since the range of
$\epsilon_{0}$ obtained from (\ref{16})  with the use of $\langle
E_{T}\rangle$ and $b$ for 1995 and 1996 Pb-Pb runs are different.
For 1995 run \cite{Abr97} we have $\epsilon_{0} \cong 2.6-2.9
GeV/fm^{-3}$ and for 1996 run \cite{Abr} $\epsilon_{0} \cong
2.0-2.1 GeV/fm^{-3}$, both estimates are for $\langle E_{T}\rangle
\geq 40 GeV$. Of course, where exactly the data points should be
placed is crucial for the valuation of the correctness of the
shape of theoretical curves.

When the additional disintegration in the nuclear matter is
included, also the magnitude of $J/\Psi$ suppression is comparable
with the data, but rather for greater $\sigma_{b}$. But note that
since we have one overall charmonium-baryon cross-section
$\sigma_{b}$, our final results underestimate the suppression (for
$\chi-,\;\psi'-baryon$ scattering the cross-section should be
greater than for $J/\Psi$).

As a final remark, we think that it is difficult to exclude
$J/\Psi$ scattering in the hot hadron gas entirely, as the reason
for the observed $J/\Psi$ suppression at this point. In our model
the most crucial parameter is the charmonium-baryon inelastic
cross-section and the final results depend on its value
substantially. Therefore it is of the greatest importance to
establish how this cross-section behaves in the hot hadron
environment. Some work has been done into this direction
\cite{KharSa,Mart,Mati}, but results presented there differ from
each other and are based on different models.

\section*{Acknowledgements} We would like to thank Dr K.Redlich for
very helpful discussions.

\end{document}